\documentclass[twocolumn]{revtex4-1}

\usepackage{graphicx}
\usepackage{epstopdf}
\usepackage{dcolumn}
\usepackage{bm}
\usepackage{amssymb}
\usepackage{hyperref}
\usepackage{url}

\begin{document}
\title{Violation of Bell-CHSH inequality using imperfect photodetectors with optical hybrid states}

\author{Hyukjoon Kwon and Hyunseok Jeong}
\affiliation{Center for Macroscopic Quantum Control, Department of Physics and Astronomy, Seoul National University, Seoul, 151-742, Korea}
\date{\today}

\begin{abstract}
We show that a Bell inequality test using an optical hybrid state between a polarized single photon and a coherent field can be highly robust against detection inefficiency.
The Bell violation occurs 
until the efficiency becomes as low as 67\%
even though its  degree becomes 
small as the detection efficiency degrades.
We consider on/off and photon number parity measurements, respectively, for the Bell test 
and they result in the similar conditions.
If the detection efficiency is higher than 98.68\%, parity measurements give
larger Bell violations close to Cirel'son's bound,
while on/off measurements give larger but moderate violations for realistic values of detector efficiency. 
Experimental realization of our proposal seems feasible in the near future for the implementation of a loophole-free Bell inequality test.
\end{abstract}

\pacs{}
\maketitle

\section{Introduction}
Einstein, Podolsky and Rosen (EPR)'s argument provoked debates upon incompatibility between quantum mechanics and local realism \cite{PhysRev.47.777}. Around thirty years after EPR's work was published, 
Bell in his celebrated paper suggested an inequality that enables one to test quantum mechanics against local realism \cite{bell1966problem}.
Since then, various versions of Bell's inequality have been suggested including
Clauser, Horne, Shimony, and Holt (CHSH)'s one 
known as the Bell-CHSH inequality
\cite{PhysRevLett.23.880}. 
The Bell-CHSH inequality has been theoretically studied within the frameworks of {\it N}-dimensional systems \cite{Gisin199215,PhysRevLett.85.4418, PhysRevA.64.024101} and continuous variables  in phase space \cite{jlk2000,Ralph2000,PhysRevLett.82.2009, PhysRevLett.88.040406,PhysRevA.67.012106,  PhysRevA.66.043803}.

Meanwhile, various experimental efforts have been made to observe violation of Bell's inequality, yet no experiment has been found to be completely loophole-free.
In general, experiments using atoms \cite{rowe2001experimental, ansmann2009violation} have suffered from the locality loophole \cite{Bell198141, Santos19951}, while optical experiments \cite{PhysRevLett.28.938, PhysRevLett.47.460, PhysRevLett.81.3563, PhysRevLett.81.5039} have not been free from the detection (or fair-sampling) loophole \cite{PhysRevD.2.1418}.
In order to close the detection loophole for the Bell-CHSH inequality test, 82.8\% of detector efficiency is required when using maximally entangled bipartite system. It was shown that Bell type inequality with non-maximally entangled states could lower the threshold efficiency to 66.7\% \cite{PhysRevA.47.R747}. 
The experimental observation of Bell inequality violation using partially entangled photon without the fair-sampling assumption was reported lately \cite{giustina2013bell}.

In order to lower the detection efficiency threshold for a loophole-free Bell test, schemes based on high dimensional \cite{PhysRevA.66.052112, PhysRevA.71.052314, collins2004relevant, PhysRevLett.98.220403, PhysRevLett.104.060401}, multiphoton \cite{PhysRevA.85.012104, PhysRevA.85.062112}, and multimode \cite{Brask2011} states have been suggested. A study on an asymmetric Bell type inequality, assuming perfect detection on one side, shows that 43\% of detection efficiency is required \cite{PhysRevLett.98.220403}, while a scheme using qudit systems requires 61.8\% of threshold efficiency \cite{PhysRevLett.104.060401}.
In principle, macroscopic entanglement enables one to perform a Bell inequality test free from the detection inefficiency
\cite{PhysRevA.85.062112}.
  Continuous variable systems with homodyne detection have also been investigated to close the detection loophole \cite{PhysRevLett.93.020401,PhysRevLett.93.130409,JiNha2010, Brask2012}.
Atom-fields entanglement to combine the advantages of both the atomic and optical systems
have been studied 
in the context of loophole-free Bell inequality tests \cite{PhysRevA.84.052122, PhysRevA.84.032102, PhysRevLett.91.110405, PhysRevA.85.022120, JiNha2010, Brask2011, Araujo2012, Teo2013} and several related experiments have been reported \cite{PhysRevLett.93.090410, PhysRevLett.96.030404, PhysRevLett.100.150404}.
It was shown that a hybrid detection scheme, combining homodyne detection and photodetection, 
with atom-photon entanglement may be used for loophole-free Bell tests under moderate transmission losses and detection efficiencies \cite{Araujo2012, Teo2013}.

In this paper, we study optical hybrid entanglement between a polarized single photon and a coherent-state field
for Bell inequality tests using inefficient detectors. We employ two different kinds of measurements for the coherent-state field, photon on/off measurement and photon number parity measurement, to investigate the Bell-CHSH inequality.
We find that the Bell-CHSH inequality is violated for low coherent amplitudes ($|\alpha|<1.0$) with detection efficiency higher than 67\%.
When realistic detection efficiency is assumed ({\it i.e.,} smaller than 98.68\%), the scheme based on on/off measurements gives larger Bell violation than the one based on photon number parity measurements, while nearly perfect detector efficiency provides higher Bell values close to  Cirel'son's bound $2\sqrt{2}$ for the parity measurement scheme. 
However, threshold values for detection efficiencies over which Bell violations occur are similar for both the measurement schemes.

\section{Bell-CHSH inequality test with optical hybrid states}
In this paper, we are interested in an optical hybrid state with entanglement between a polarized single photon and a coherent state,
\begin{equation}
\left| \Psi \right> = \frac{1}{\sqrt{2}} \left( {\left| H \right>}_{A} {\left| \alpha \right>}_{B} + {\left| V \right>}_{A}{\left| -\alpha \right>}_{B} \right)
\label{HybridState}
\end{equation}
where $\left|H\right>$ and $\left|V\right>$ refer to horizontal and vertical polarization state of a photon each, and $\left|\pm \alpha \right>$ are coherent states of amplitudes $\pm\alpha$.  
Such hybrid entangled states are particularly useful for deterministic quantum teleportation and resource-efficient quantum computing using linear optics \cite{PhysRevA.87.022326}  as well as for information transfer between different types of qubits \cite{Park2012}.
It was shown that
this type of entanglement can be obtained if a weak cross-Kerr nonlinear interaction is available
\cite{Jeong2005weak}. It is highly challenging to implement cross-Kerr nonlinearity with high-fidelity \cite{PhysRevA.73.062305, Shapiro2007}, while there have been several proposals to obtain such high fidelity cross-Kerr interactions \cite{PhysRevA.83.053826, mahdi2012memory, PhysRevA.87.042325}.

\subsection{Bell-CHSH inequality using on/off and parity measurements}
In order to perform Bell inequality tests, an entangled state should be shared by two locally separate parties. With regard to the state in Eq.~(\ref{HybridState}), the single photon part with the polarization degree of freedom and the coherent state part with amplitudes $\pm\alpha$ are subscripted by {\it A} and {\it B}, respectively. Each party may locally perform unitary operations and dichotomic measurements.
In order to construct a  Bell-CHSH inequality, each measurement outcome is determined as either $+1$ or $-1$. We may choose $\hat{\Pi}_{A} = \left| H \right> \left< H \right| - \left| V \right> \left< V \right| $ for polarization measurements for the single-photon part and
\small
\begin{equation}
\hat{\Pi}_{B} = 
\left\{
\begin{array}{cc}
\left| 0 \right> \left< 0 \right| - \displaystyle\sum_{n=1}^{\infty} {\left| n \right> \left< n \right|} & \left( \mathrm{on/off} \right) \\
\displaystyle\sum_{n=0}^{\infty} {\left( \left| 2n \right> \left< 2n \right| - \left| 2n+1 \right> \left< 2n+1 \right| \right) } & \left( \mathrm{parity} \right)\\
\end{array}
\right.
\end{equation}
\normalsize
for on/off and photon number parity measurements each.
Outcomes $\pm1$ denote no-click/click events for on/off measurements and even/odd number results for photon number parity measurements.

An arbitrary unitary operation on a single photon qubit with the qubit basis of $|H\rangle$ and $|V\rangle$ can be represented by
\[
U\left( \xi \right) = 
\left(
\begin{array}{cc}
\cos \left| \xi \right| & \frac{\xi}{\left| \xi \right|} \sin \left| \xi \right| \\
-\frac{\xi^{*}}{\left| \xi \right|} \sin\left| \xi \right| & \cos\left| \xi \right| \\
\end{array}
\right)
\]
with complex variable $\xi$.
The displacement operation, 
$D \left( \beta \right) = e^{\beta \hat{a}^{\dagger} - \beta^{*} \hat{a}}$,
is used a unitary operation on the coherent state part ({\it i.e.}, mode $B$),
where  $\beta$ is a complex variable.
A previous result shows that the displacement operator approximately
acts as a qubit rotation for a coherent-state qubit with basis $|\pm\alpha\rangle$ \cite{PhysRevA.67.012106}.
The expectation value of the joint measurement is obtained as
\begin{equation}
	E \left( \xi, \beta \right) = \left< \hat{O}_{A} \otimes \hat{O}_{B} \right>, \\
\end{equation}
where 
$\hat{O}_{A}\left( \xi \right) = U\left( \xi \right) \hat{\Pi}_{A}U^{\dagger}\left( \xi \right)$
and $\hat{O}_{B}\left( \beta \right) = D\left( \beta \right) \hat{\Pi}_{B}D^{\dagger} \left( \beta \right)$.
The Bell-CHSH inequality  is then defined as
$\left| B \left( \xi_{1}, \xi_{2}, \beta_{1}, \beta_{2} \right) \right| \leq 2$
with the Bell function 
\begin{equation}
\label B
\begin{array}{l}
  B \left( \xi_{1}, \xi_{2}, \beta_{1}, \beta_{2} \right) \\
 \quad = E \left( \xi_{1}, \beta_{1} \right)  + E \left( \xi_{1}, \beta_{2} \right) + E \left( \xi_{2}, \beta_{2} \right) - E \left( \xi_{2}, \beta_{1} \right).
\label{Bellneq}
\end{array}
\end{equation}
We define $\xi = - ({\theta}/{2}) e^{-i\phi}$ and $\beta = \left| \beta \right| e^{i\Phi}$ with $ 0 \leq \theta < \pi$, $ 0 \leq \phi$ and $\Phi < 2\pi$ for simplicity. Without loss of generality, we take $\alpha$ to be real  because the phase of $\alpha$ may absorbed by $\Phi$.
We obtain the expectation values as
\small
\begin{equation}
\begin{array}{l}
	E^{\mathrm{On/off}}\left(\theta,\phi,\left| \beta \right|, \Phi \right) \\
\qquad\qquad = 2\cos\theta e^{-\left( {\left| \alpha \right|}^2 + {\left| \beta \right|}^2 \right)}\sinh\left(2\left|\alpha\right|\left|\beta\right|\cos\Phi\right) \\
\qquad\qquad\qquad\qquad + 2\sin\theta e^{-\left( {\left| \alpha \right|}^2 + {\left| \beta \right|}^2 \right)} \cos\left(2\left|\alpha\right|\left|\beta\right|\sin\Phi - \phi \right) \\
\qquad\qquad\qquad\qquad\qquad - \sin\theta e^{-2{\left|\alpha\right|}^2} \cos\phi,
\end{array}
\end{equation}
\normalsize
\small
\begin{equation}
\begin{array}{c}
	E^{\mathrm{Parity}}\left(\theta,\phi,\left| \beta \right|, \Phi \right) = \cos\theta e^{-2 \left( {\left| \alpha \right|}^2 + {\left| \beta \right|}^2 \right)}\sinh\left(4\left|\alpha\right|\left|\beta\right|\cos\Phi\right) \\
\qquad\qquad\qquad\qquad\qquad + \sin \theta e^{-2 {\left| \beta \right|}^2} \cos\left(4\left|\alpha\right|\left|\beta\right|\sin\Phi - \phi \right)
\end{array}
\end{equation}
\normalsize
by applying on/off and photon number parity measurements, respectively,  on the coherent-state part.

\subsection{Photodetector efficiency and the detection loophole}
A physical model of an imperfect photodetector with detection efficiency $p$ is described by a beam splitter of transmission coefficient $\sqrt{p}$ before a perfect photodetector.
In terms of positive operator valued measurement (POVM) with the photon number basis,
a  photodetector with efficiency $p$ may be written as \cite{kok2010introduction},
\begin{equation}
	\hat{E}_p^{\left(n\right)} = \displaystyle\sum_{m=0}^{\infty}
\left(\!
\begin{array}{c}
n+m\\
n\\
\end{array}
\!\right)
p^n {\left(1-p\right)}^{m} \left| n+m \right> \left< n+m \right|.
\end{equation}
Then the effective on/off measurement of detection efficiency $\eta_B$ becomes
\small
\begin{eqnarray}
\begin{array}{l}
\hat{\Pi}_{B,{\mathrm{eff}}}^{\mathrm{On/off}}  = \hat{E}_{\eta_B}^{\left(0\right)} - \displaystyle\sum_{n=1}^{\infty}\hat{E}_{\eta_B}^{\left(n\right)} \\
 \quad \quad	 = \displaystyle\sum_{m=0}^{\infty} \left[
 2 {\left( 1 - \eta_B \right)}^m \left| m \right> \left< m \right|
\begin{array}{c}
\\\\
\end{array}\right. \\
	  \quad\qquad 
- \left. \displaystyle\sum_{n=0}^{\infty}
\left(\!
\begin{array}{c}
n+m\\
n\\
\end{array}
\!\right)
\eta_B^{n} {\left( 1- \eta_B \right)}^m \left| n+m \right> \left< n+m \right| \right]. \\
\end{array}
\end{eqnarray}
\normalsize
Similarly, the effective photon number parity measurement is given by
\small
\begin{equation}
\label{eq:e_eff}
\begin{array}{l}
\hat{\Pi}_{B,{\mathrm{eff}}}^{\mathrm{parity}} = \hat{E}_{\eta_B}^{\left(\mathrm{even}\right)} -\hat{E}_{\eta_B}^{\left(\mathrm{odd}\right)} \\
 = \displaystyle\sum_{m=0}^{\infty}\displaystyle\sum_{n=0}^{\infty} \left[ 
\left(\!
\begin{array}{c}
2n+m\\
2n\\
\end{array}
\!\right)
\eta_B^{2n} {\left( 1 - \eta_B \right)}^m \left| 2n+m \right> \left< 2n+m \right| \right. \\
- \left. 
\left(\!
\begin{array}{c}
2n+1+m\\
2n+1\\
\end{array}
\!\right)
\eta_B^{2n+1} {\left( 1- \eta_B \right)}^m \left| 2n+1+m \right> \left< 2n+1+m \right| \right]. \\
\end{array}
\end{equation}
\normalsize

In order to avoid the detection loophole, we assign $+1$ for a ``no-detection'' outcome on the polarization part. Provided the polarization measurement detection efficiency is $\eta_A$, the expectation value for the combined measurement is given by
\begin{equation}
\label{Eeff}
	E_{\mathrm{eff}} = \eta_A \left< \hat{O}_{A} \otimes \hat{O}_{B,\mathrm{eff}} \right> + \left( 1 - \eta_A \right) \mathrm{Tr}_{B} \left[ \hat{O}_{B,\mathrm{eff}} \rho_{B} \right],
\label{Eeff}
\end{equation}
where $\rho_{B}$ is a reduced density matrix obtained by tracing over polarization part $A$, {\it i.e.}
\begin{equation}
\rho_{B} = \mathrm{Tr}_{A}\left( \rho \right) = \frac{1}{2} \left( \left| \alpha \right> \left< \alpha \right| + \left| -\alpha \right> \left< -\alpha \right| \right).
\end{equation}
Each term in Eq.~(\ref{eq:e_eff}) could be directly calculated as
\small
\begin{equation}
\begin{array}{r}
\label{Oeff1}
 \left< \hat{O}_{A}\otimes \hat{O}_{B,\mathrm{eff}}^{\mathrm{On/off}} \right> = 2 \cos\theta e^{-\eta_B\left( {\left| \alpha \right|}^2 + {\left| \beta \right|}^2 \right)}\sinh\left(2\eta_B\left|\alpha\right|\left|\beta\right|\cos\Phi\right) \\
+ 2\sin\theta e^{-\left(2-\eta_B\right) {\left| \alpha \right|}^2 - \eta_B{\left| \beta \right|}^2} \cos\left(2 \eta_B \left|\alpha\right|\left|\beta\right|\sin\Phi - \phi \right) \\
- \sin\theta e^{-2{\left|\alpha\right|}^2} \cos\phi ,
\end{array}
\end{equation}
\begin{equation}
\label{Oeff2}
\mathrm{Tr}_{B} \left[ \hat{O}_{B,\mathrm{eff}}^{\mathrm{On/off}} \rho_{B} \right] = 2e^{-\eta_B \left( |\alpha|^2 + |\beta|^2 \right)} \cosh \left( 2 \eta_B|\alpha||\beta| \cos\Phi \right) -1
\end{equation}
\normalsize
for photon on/off measurements and 
\small
\begin{equation}
\begin{array}{r}
 \left< \hat{O}_{A} \hat{O}_{B,\mathrm{eff}}^{\mathrm{Parity}} \right> = \cos\theta e^{-2 \eta_B \left( {\left| \alpha \right|}^2 + {\left| \beta \right|}^2 \right)}\sinh\left(4 \eta_B \left|\alpha\right|\left|\beta\right|\cos\Phi\right) \\
+ \sin \theta e^{-2\left(1- \eta_B \right)|\alpha|^2-2\eta_B {\left| \beta \right|}^2} \cos\left(4 \eta_B \left|\alpha\right|\left|\beta\right|\sin\Phi - \phi \right), 
\label{Peff1}
\end{array}
\end{equation}
\begin{equation}
\mathrm{Tr}_{B} \left[ \hat{O}_{B,\mathrm{eff}}^{\mathrm{Parity}} \rho_{B} \right] = e^{-2 \eta_B \left( |\alpha|^2 + |\beta|^2 \right)} \cosh \left( 4 \eta_B |\alpha||\beta| \cos\Phi \right)
\label{Peff2}
\end{equation}
\normalsize
for photon number parity measurements.

\section{Optimization}

In order to observe violation of the Bell-CHSH inequality, it is important to find optimizing conditions 
for local unitary variables under which the Bell functions have largest values.
The optimizing conditions presented throughout this paper are numerically found \cite{numerical}, and wherever possible, we try to find corresponding analytical expressions.

\subsection{Perfect photodetector efficiency}
\begin{figure}[b]
	\centering
	\includegraphics[width=7.cm]{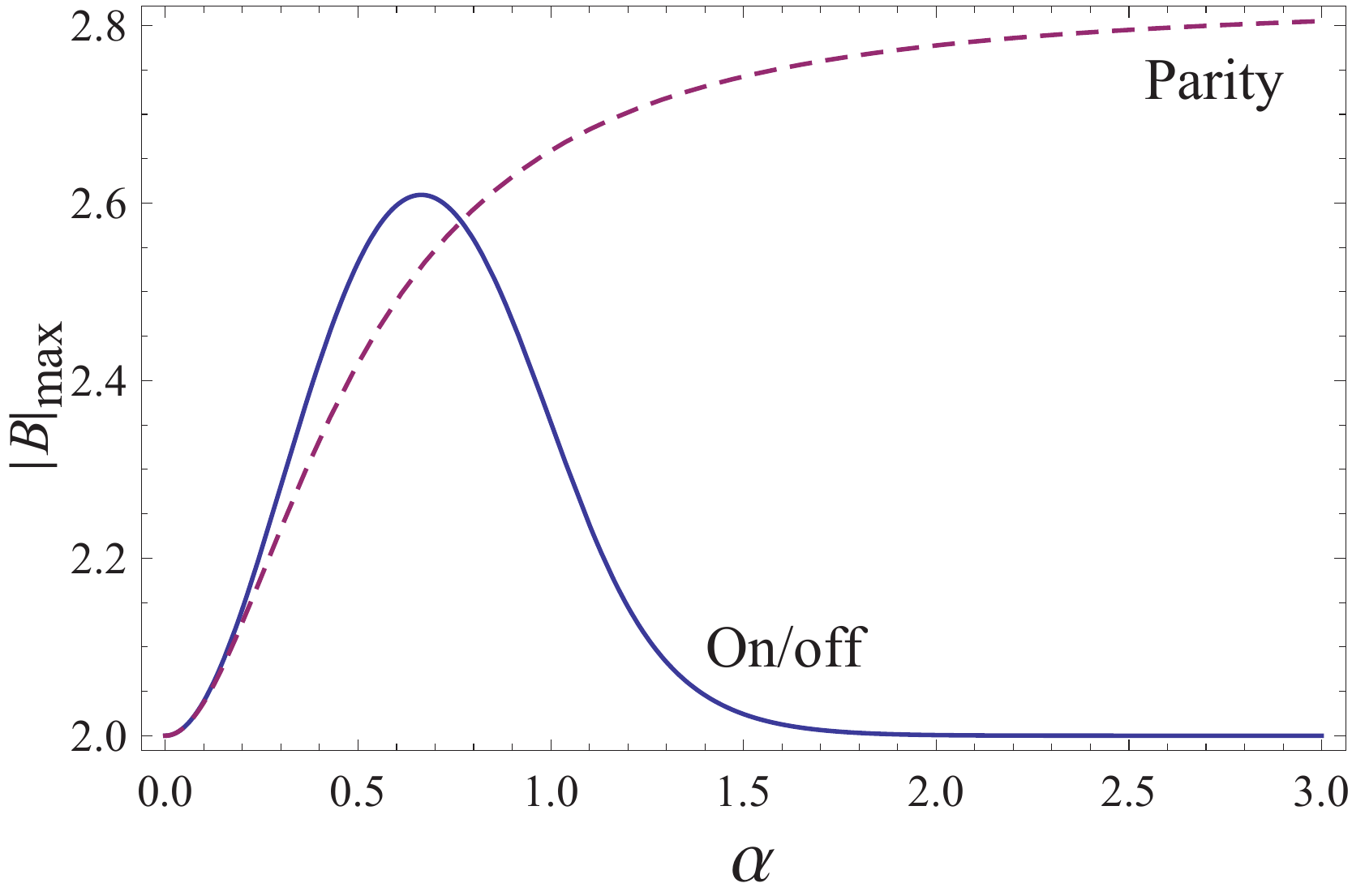}
	\caption{\small (Color online) Maximized Bell value $|B|_{\rm max}$ for varying $\alpha$ with perfect detector efficiency. The solid curve refers to photon on/off measurements while the dashed curve to photon number parity measurements.}
\label{fig:PDB}
\end{figure}

We first suppose perfect efficiencies for all detectors used for Bell inequality tests.
In the case of photon on/off measurements, optimizing conditions can be obtained as
\cite{PhysRevA.85.022120}
\begin{equation}
\xi_1 = -\frac{\pi}{4},~ \xi_2 = 0,~ \beta_1 = -\beta_2 = -|\beta|
\end{equation}
with $|\beta|$ satisfying
\begin{equation}
|\beta|\left(1+\sinh\left(2|\alpha||\beta|\right)\right) = |\alpha| \cosh\left(2|\alpha||\beta|\right).
\end{equation}
We plot the Bell function in Fig.~\ref{fig:PDB} and note that as the coherent amplitude $|\alpha|$ increases the maximized Bell value increases up to $\left| B \right| ^{\mathrm{On/off}}_\mathrm{max} \approx 2.61$ for $|\alpha|\approx0.664$. However, Fig.~\ref{fig:PDB} also shows that further increase of $|\alpha|$ results in lower maximized Bell values. This 
can be attributed to the fact that the probability of ``no-click'' on a photodetector becomes lower
when $|\alpha|$ becomes larger \cite{PhysRevA.67.012106}.

We also find the optimizing conditions for photon number parity measurements as
\begin{equation}
\xi_1 = -\frac{\pi}{4},~\xi_2 = i\frac{\pi}{4},~\beta_1 = -\beta_2 = -i|\beta|
\end{equation}
with 
$|\beta|$ satisfying
$\tan\left[4|\alpha||\beta|\right] = (|\alpha|-|\beta|)/(|\alpha|+|\beta|)$
nearest to zero. One may expect that Bell value would increase as $|\alpha|$ increases because probabilities to have even and odd photon number in a coherent state become equal as $|\alpha|\rightarrow\infty$. In practice, we note that Bell value of parity measurements rapidly approaches to Cirel'son's bound $2\sqrt{2}$ when $|\alpha|\gg 1$.

\subsection{Imperfect detection for coherent-state fields}
\begin{figure}
	\centering
	\includegraphics[width=4.2cm]{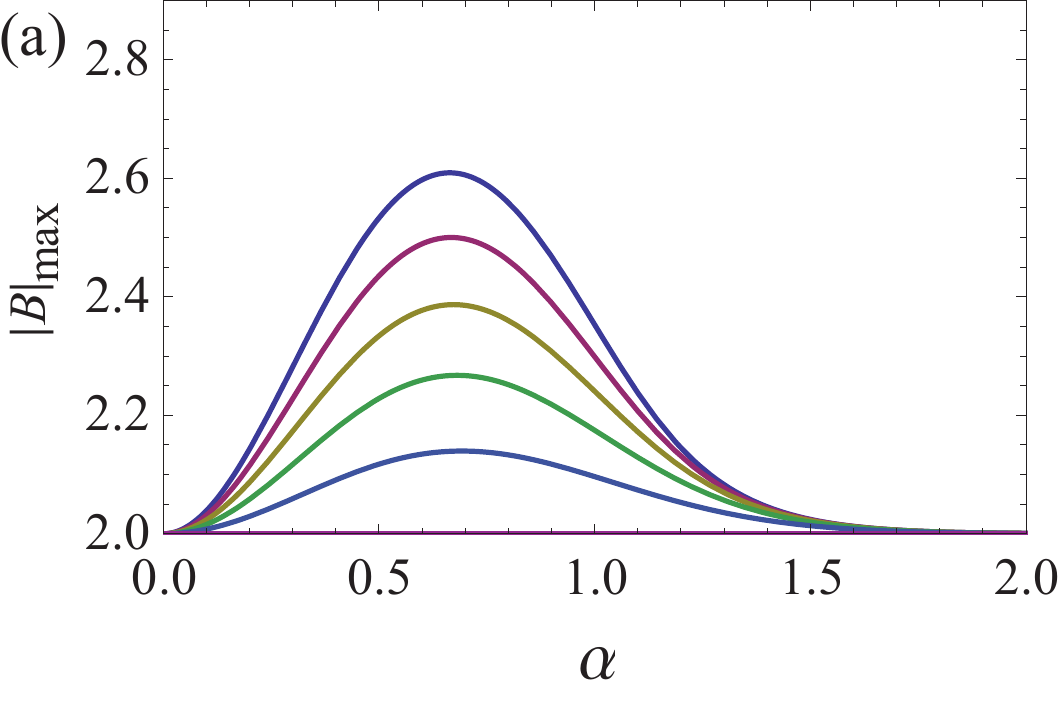}
	\includegraphics[width=4.2cm]{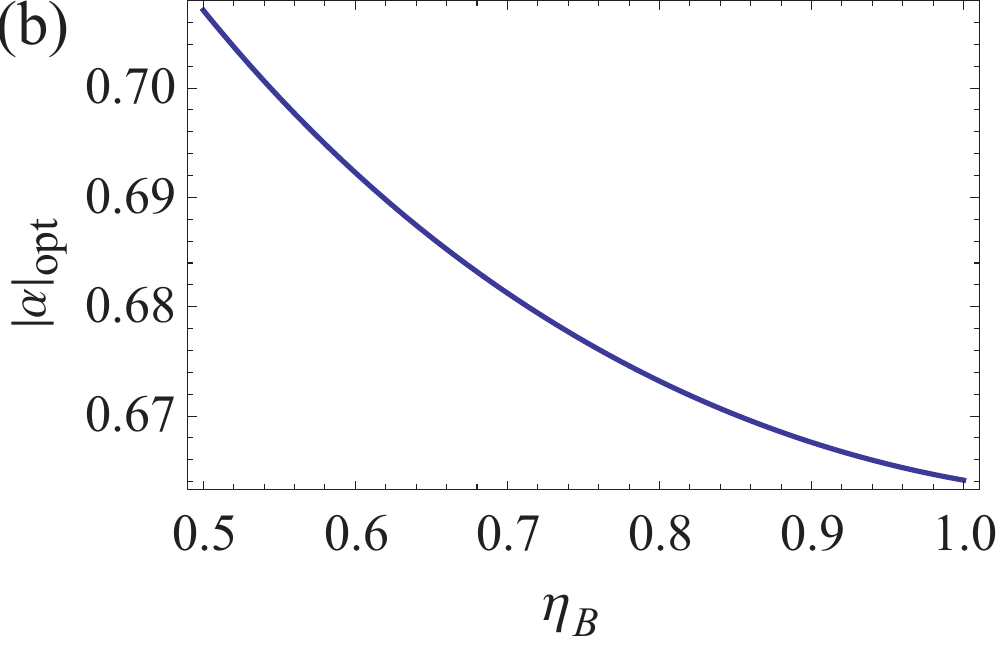} \\
	\includegraphics[width=4.2cm]{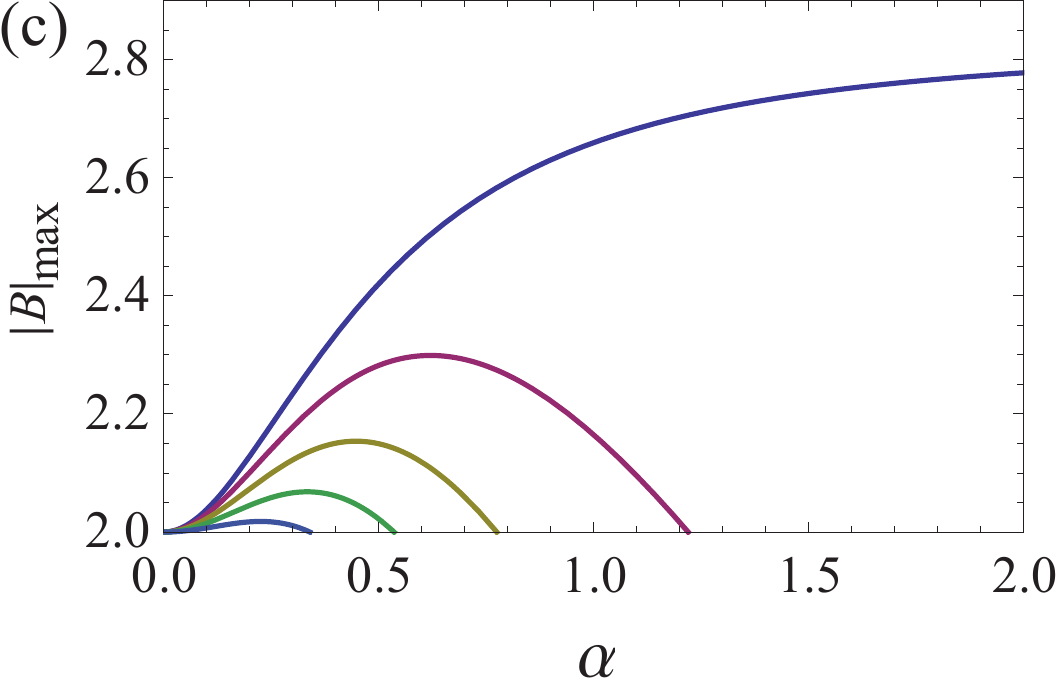} 
	\includegraphics[width=4.2cm]{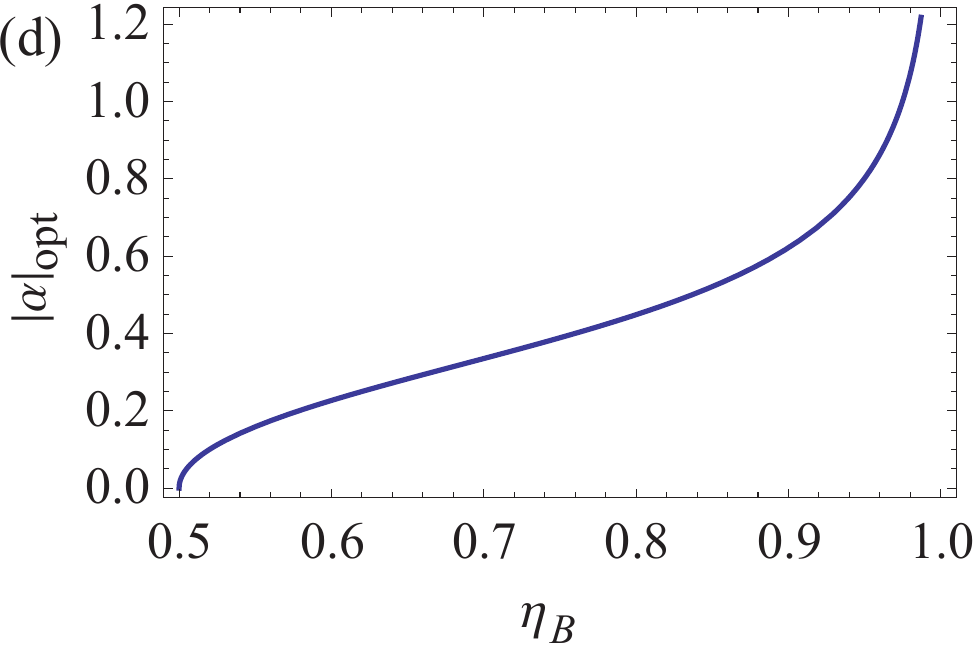} 
	\caption{\small (Color online) (a) Maximized Bell value for on/off measurements $|B|_\mathrm{max}^\mathrm{on/off}$  and (b) optimizing $|\alpha|_\mathrm{opt}$ for each coherent measurement detection efficiency. (c) Maximized Bell value for parity measurements $|B|_\mathrm{max}^\mathrm{parity}$ and (d) optimizing $|\alpha|_\mathrm{opt}$. In both the cases, the detection efficiency $\eta_B$ for the coherent-state part decreases by 0.1 from the perfect value  ($\eta_B=1$) to $\eta_B=0.5$.}
	\label{Coherentp}
\end{figure}
Now we consider the situation of perfect polarization measurements $\left(\eta_A=1\right)$ and imperfect coherent field measurements $\left( \eta_B \neq 1 \right)$.
Optimizing conditions for photon on/off measurements can be obtained by $\xi_1 = -\pi/4$, $\xi_2 = 0$ and $\beta_1 = -\beta_2 = -|\beta|$ to be real with $|\beta|$ satisfying
\small
\begin{equation}
	|\beta| e^{-2(1-\eta_B)|\alpha|^2} + |\beta| \sinh \left( 2 \eta_B |\alpha||\beta| \right)  - |\alpha|\cosh\left(2 \eta_B |\alpha||\beta|\right) = 0.
\end{equation}
\normalsize
In this case, the Bell value becomes
\small
\begin{equation}
\begin{array}{l}
B^{\mathrm{On/off}} \left( \xi_1, \xi_2, \beta_1, \beta_2 \right) \\
= 4 e^{-\eta_B \left(|\alpha|^2 +|\beta|^2 \right)} \left[ e^{-\left( 1-\eta_B \right) |\alpha|^2 } + \sinh \left( 2 \eta_B |\alpha||\beta|\right)\right] -2e^{-2|\alpha|^2} .
\end{array}
\end{equation}
\normalsize
Figure~\ref{Coherentp} shows that the maximum Bell value is obtained for $0.66 < |\alpha| < 0.71$ and the optimizing coherent amplitude $|\alpha_\mathrm{opt}|$ monotonically decreases as the detector efficiency increases. Violation of the Bell-CHSH inequality occurs until the detection efficiency reaches 0.5. This result is consistent with the equivalent Bell inequality test using entanglement between an atom and a coherent state in a cavity \cite{PhysRevA.85.022120}.

We find that optimizing conditions for photon number parity measurements
are $\xi_1 = -{\pi}/{4}$, $\xi_2 = {i\pi}/{4}$ and $\beta_1 = - \beta_2 = -i |\beta|$ to be pure imaginary. Here, $|\beta|$ is the solution of
\begin{equation}
	\tan \left( 4 \eta_B |\alpha||\beta| \right) = \frac{|\alpha|-|\beta|}{|\alpha|+|\beta|}
\end{equation}
nearest to zero, and the Bell function is
\small
\begin{equation}
\begin{array}{l}
B^{\mathrm{Parity}} \left( \xi_1, \xi_2, \beta_1, \beta_2 \right) \\
\quad = 2 e^{-2(1-\eta_B)|\alpha|^2 - 2 \eta_B |\beta|^2} \left( \cos\left(4 \eta_B |\alpha||\beta|\right) + \sin\left(4 \eta_B |\alpha||\beta|\right)\right).
\end{array}
\end{equation}
\normalsize
The optimizing coherent amplitude, $|\alpha_\mathrm{opt}|$, increases when
the detection efficiency becomes larger.
This is opposite to the case on/off measurement scheme, due to the fact when the efficiency of the photon number parity measurement is low, $|\alpha|$ should be small to reduce the possibility of parity flips.
In most of imperfect detector efficiency conditions, the on/off measurement scheme gives higher violation of the Bell-CHSH inequality than the parity measurement scheme. However, the values of the detection efficiency required  to violate the Bell-CHSH inequality are the same (50\%) for both the schemes.

\subsection{Imperfect detectors for both measurements}
We now consider the most realistic case in which both the polarization measurement and the coherent field measurement are imperfect ($\eta_A<1$ 
and $\eta_B<1$). In this case, a nontrivial calculation is needed to obtain the optimizing conditions. It is still sufficient to take real $\xi$ and $\beta$ for optimizing conditions of photon on/off measurements, but with $|\beta_1| \neq |\beta_2|$. 
On the other hand, the optimizing parameters for photon number parity measurements tend to have different conditions by detection efficiency of photodetector.
When the detection efficiency is high, we find $|\xi_1| = |\xi_2| = {\pi}/{4}$ and $\beta_1,\beta_2$ to be pure imaginary for the optimization.
If the detector efficiency is low, the optimizing conditions could be chosen to be the same with those of the on/off measurement scheme (see Appendix).

\begin{figure}[t]
	\centering
	\includegraphics[width=4.2cm]{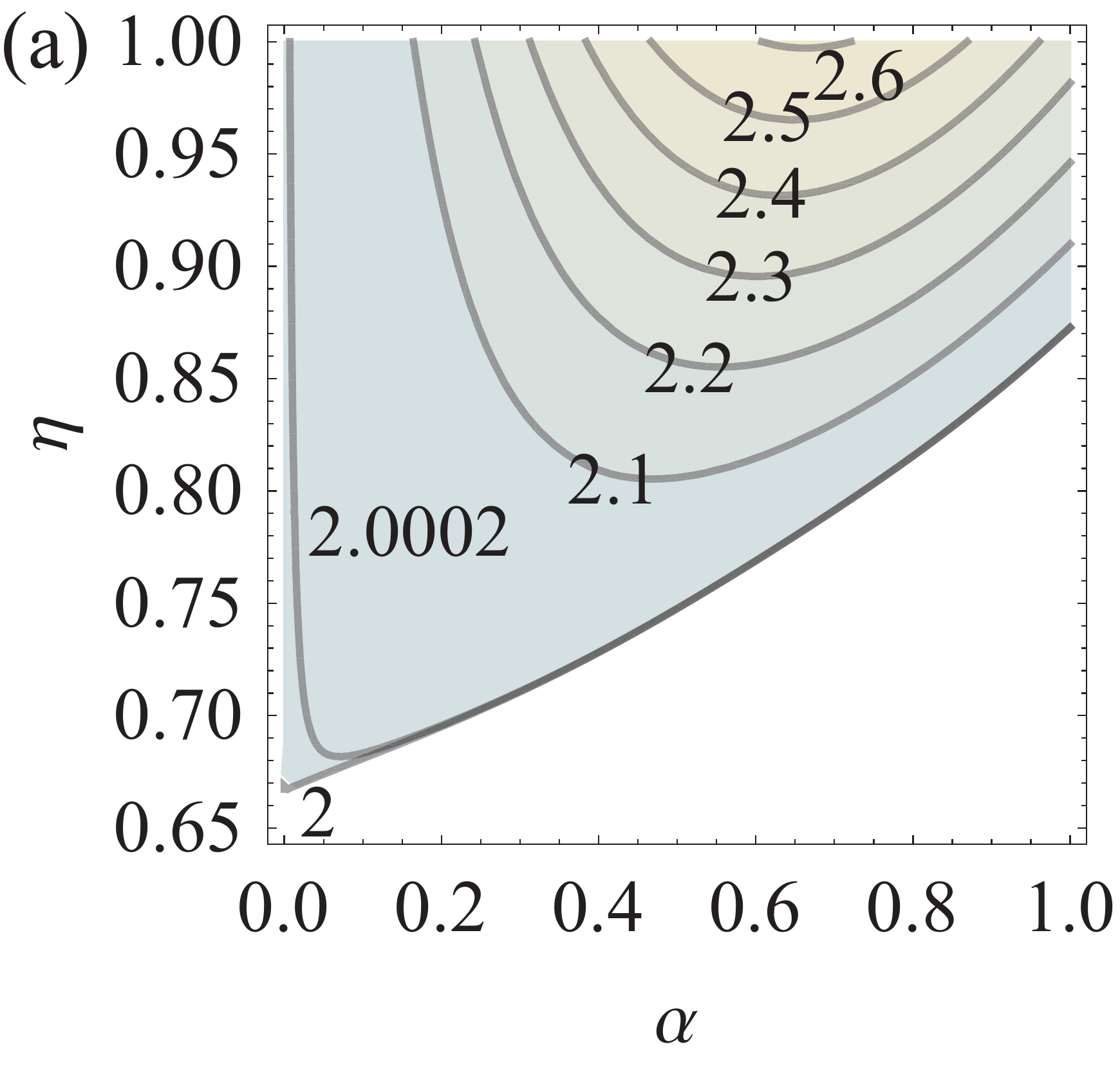}
	\includegraphics[width=4.2cm]{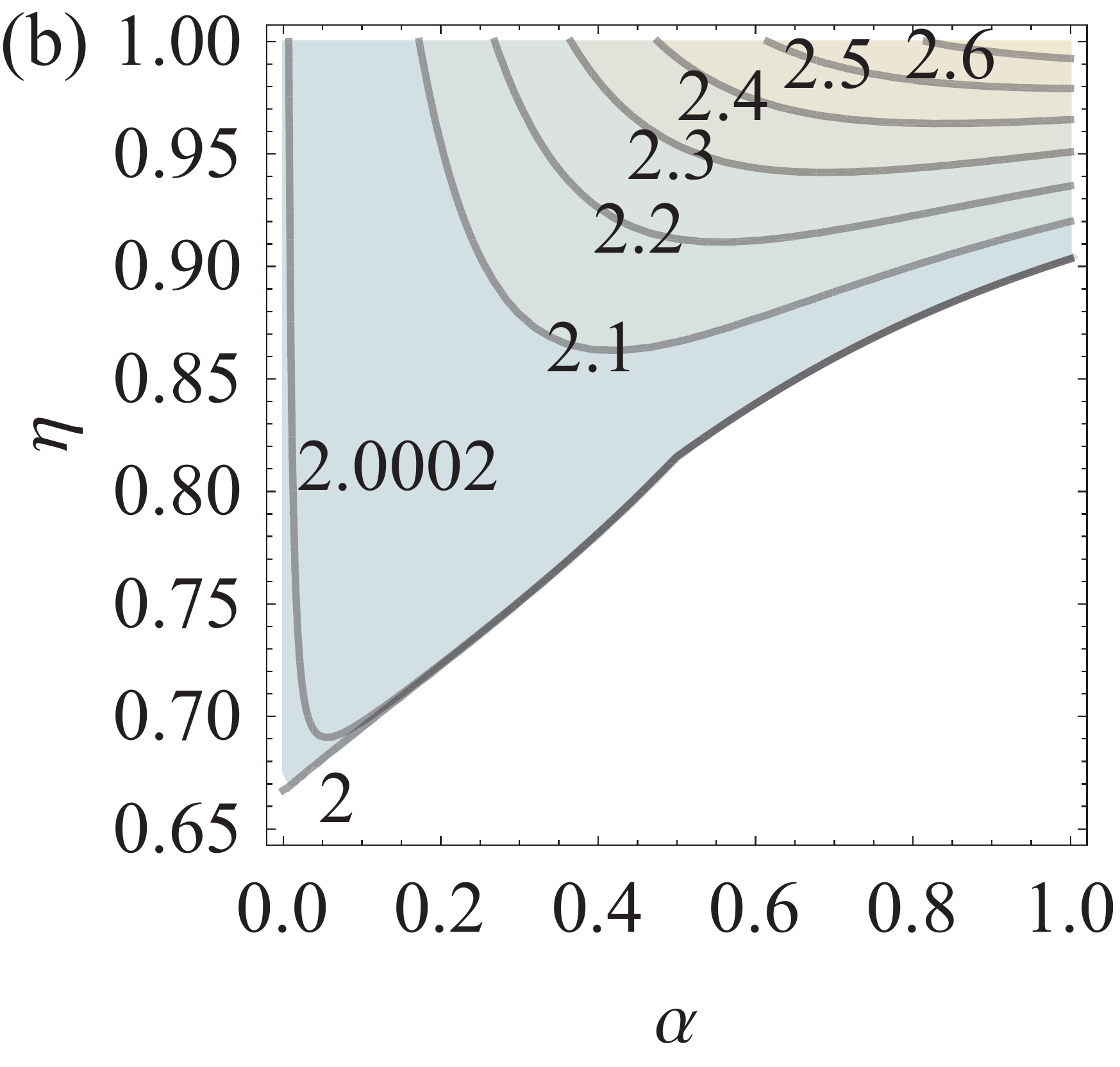}
	\caption{\small (Color online) Maximized Bell-CHSH functions in terms of  detection efficiency $\eta$ for both modes and coherent amplitude $\alpha$ for (a) on/off and (b) parity measurement schemes. The detection efficiency threshold to violate the Bell-CHSH inequality is about 67\% when the coherent amplitude is low.}
	\label{Contourna}
\end{figure}
\begin{figure}[b]
	\centering
	\includegraphics[width=4.2cm]{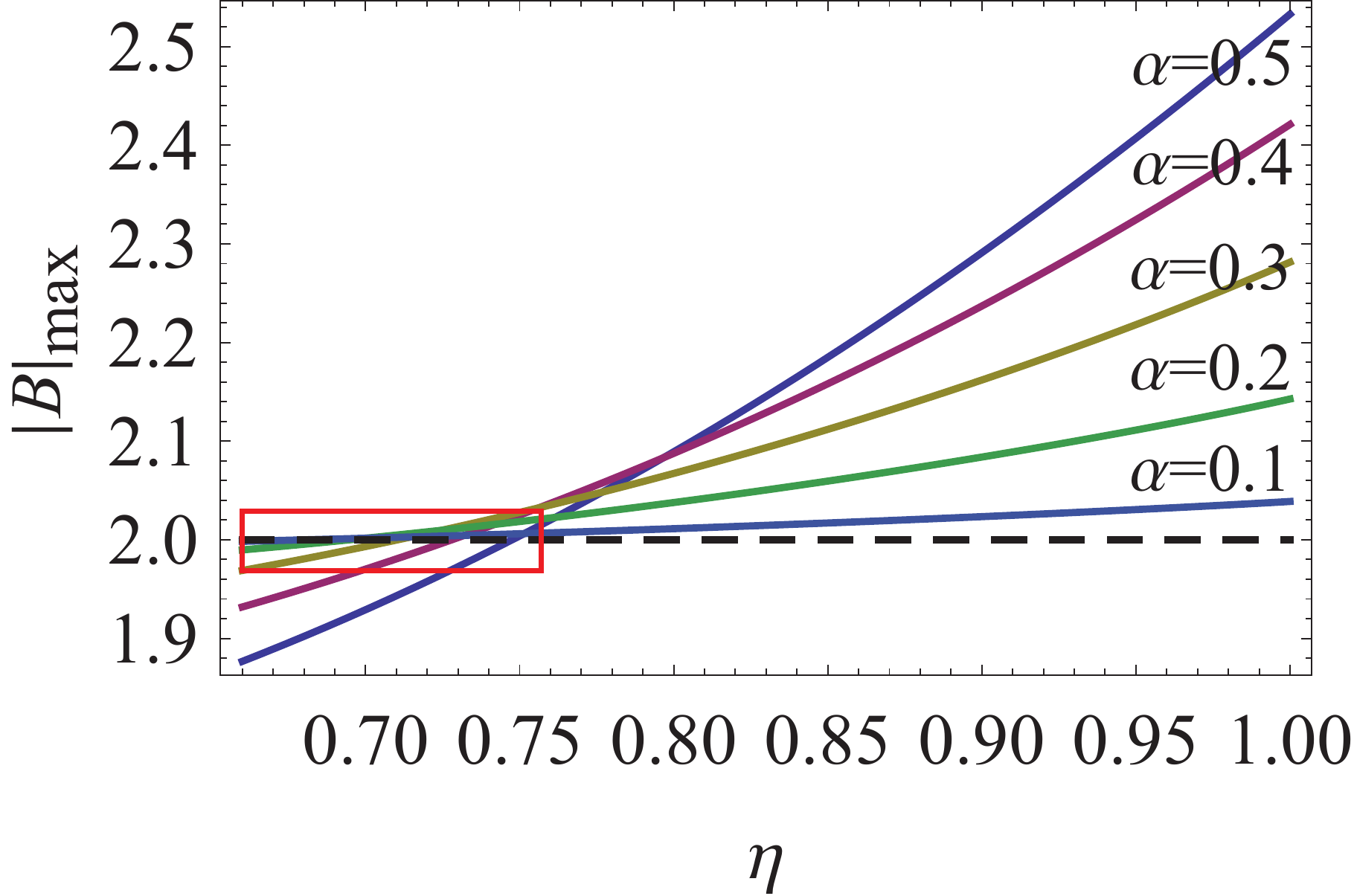}	
	\includegraphics[width=4.2cm]{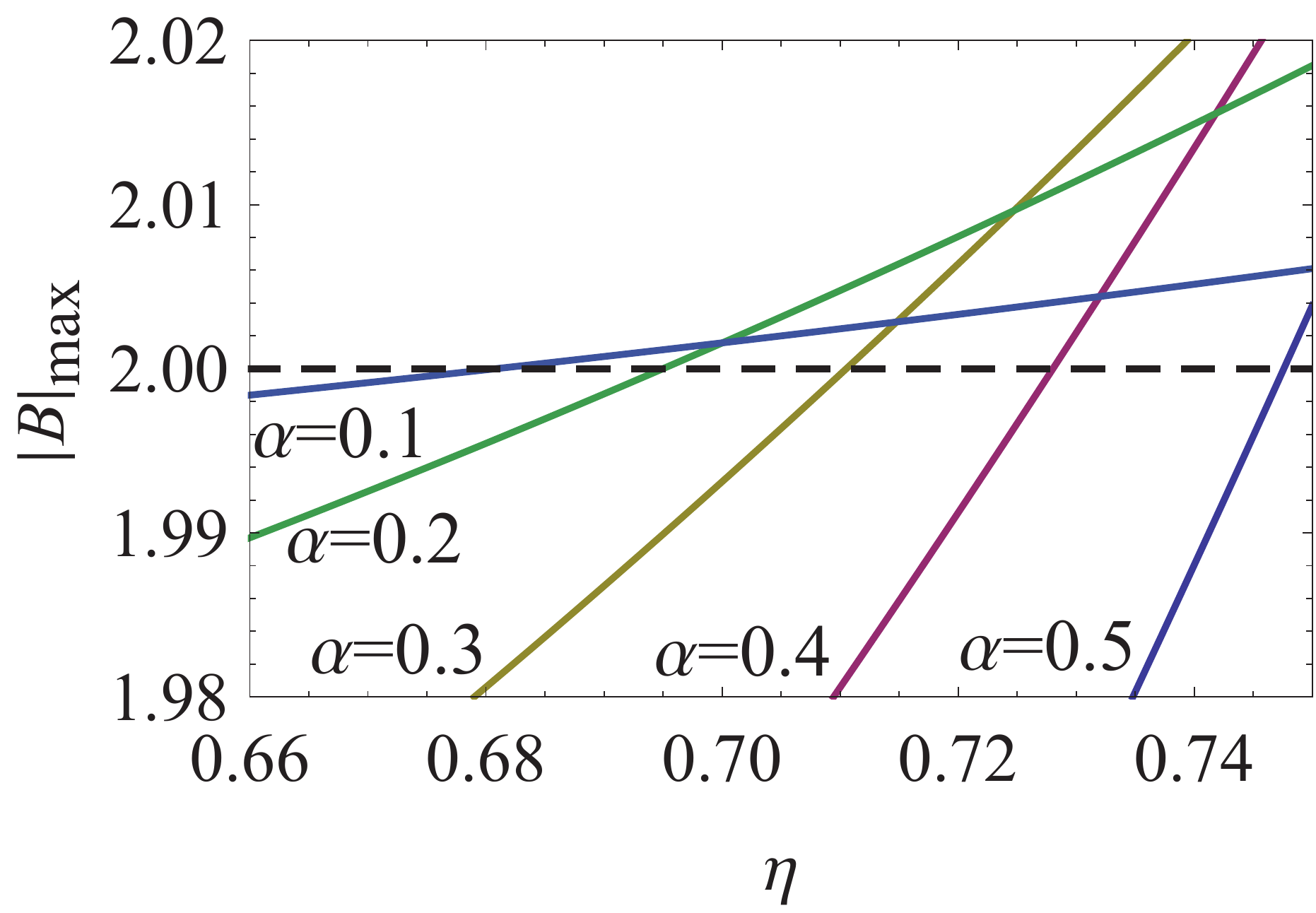}	
	\caption{\small (Color online) Maximum Bell value via detector efficiency with varying coherent amplitude from 0.1 to 0.5 for the on/off measurement scheme. The right-hand-side figure represents the boxed region of the left-hand-side one.}
\label{coherentscaleO}
\end{figure}
\begin{figure}[t]
	\centering
	\includegraphics[width=4.25cm]{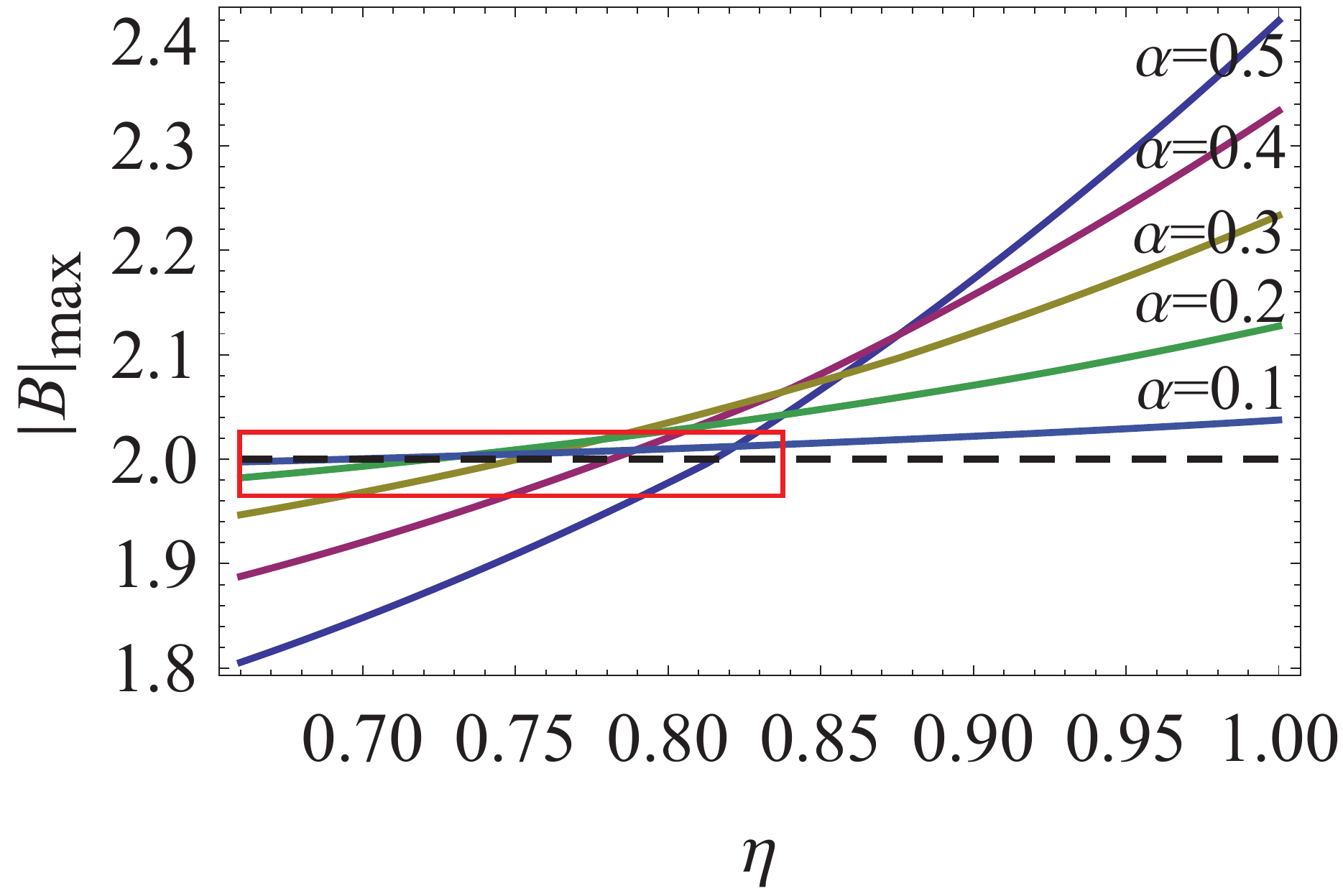}	
	\includegraphics[width=4.15cm]{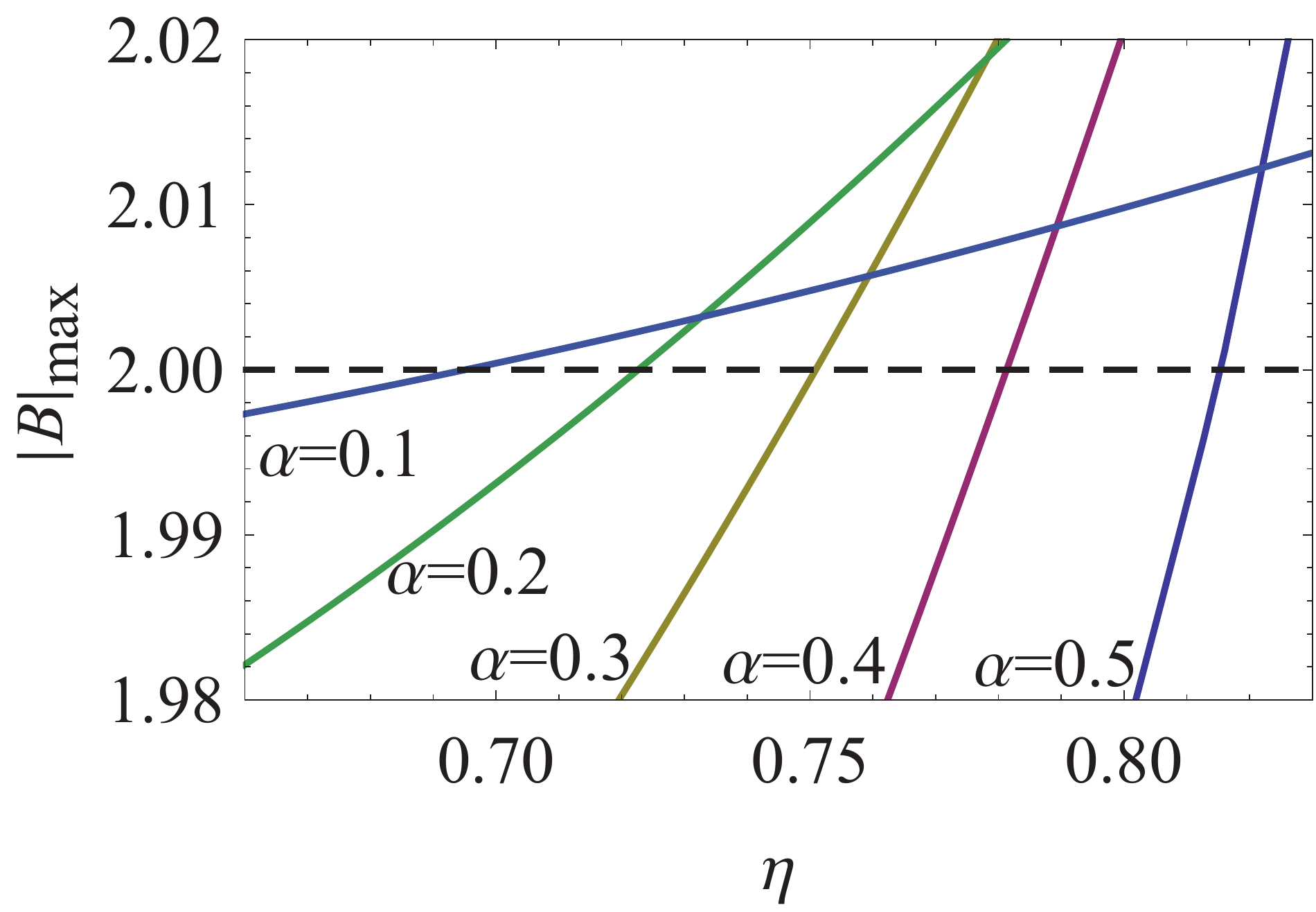}	
	\caption{\small (Color online) Maximum Bell values against detection efficiency with varying coherent amplitudes from 0.1 to 0.5 for the parity measurement scheme. The right-hand-side figure represents the boxed region of the left-hand-side one.}
\label{coherentscaleP}
\end{figure}

\begin{figure}[b]
	\centering
	\includegraphics[width=4.1cm]{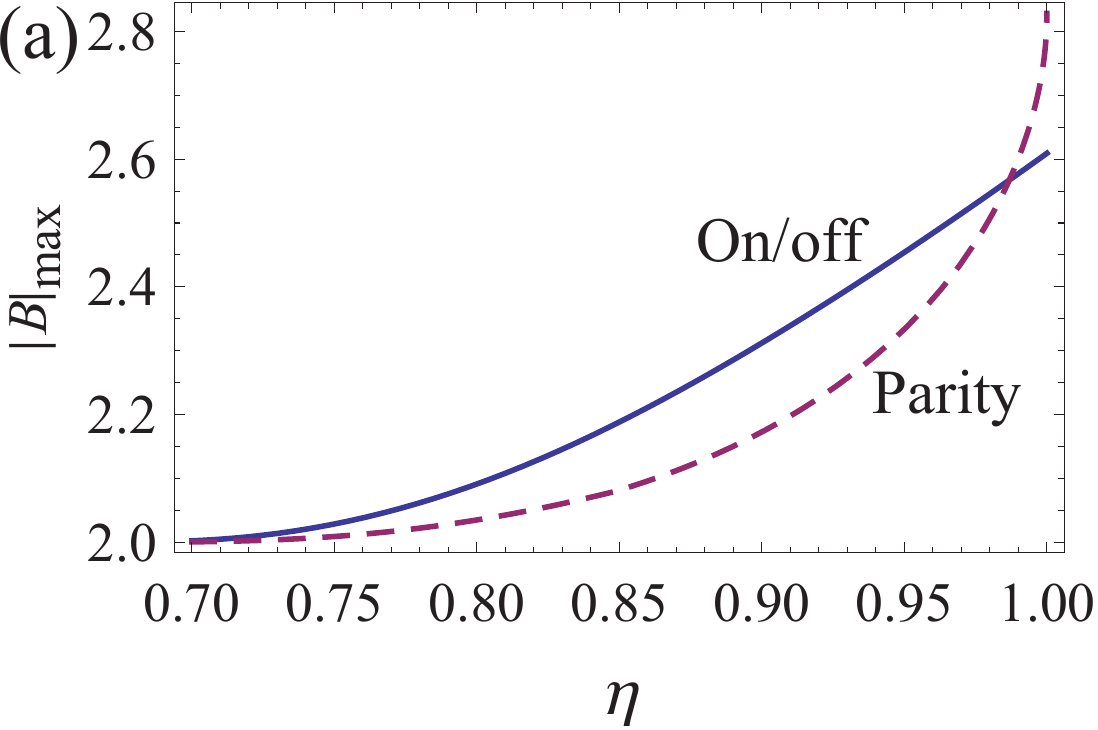}
	\includegraphics[width=4.25cm]{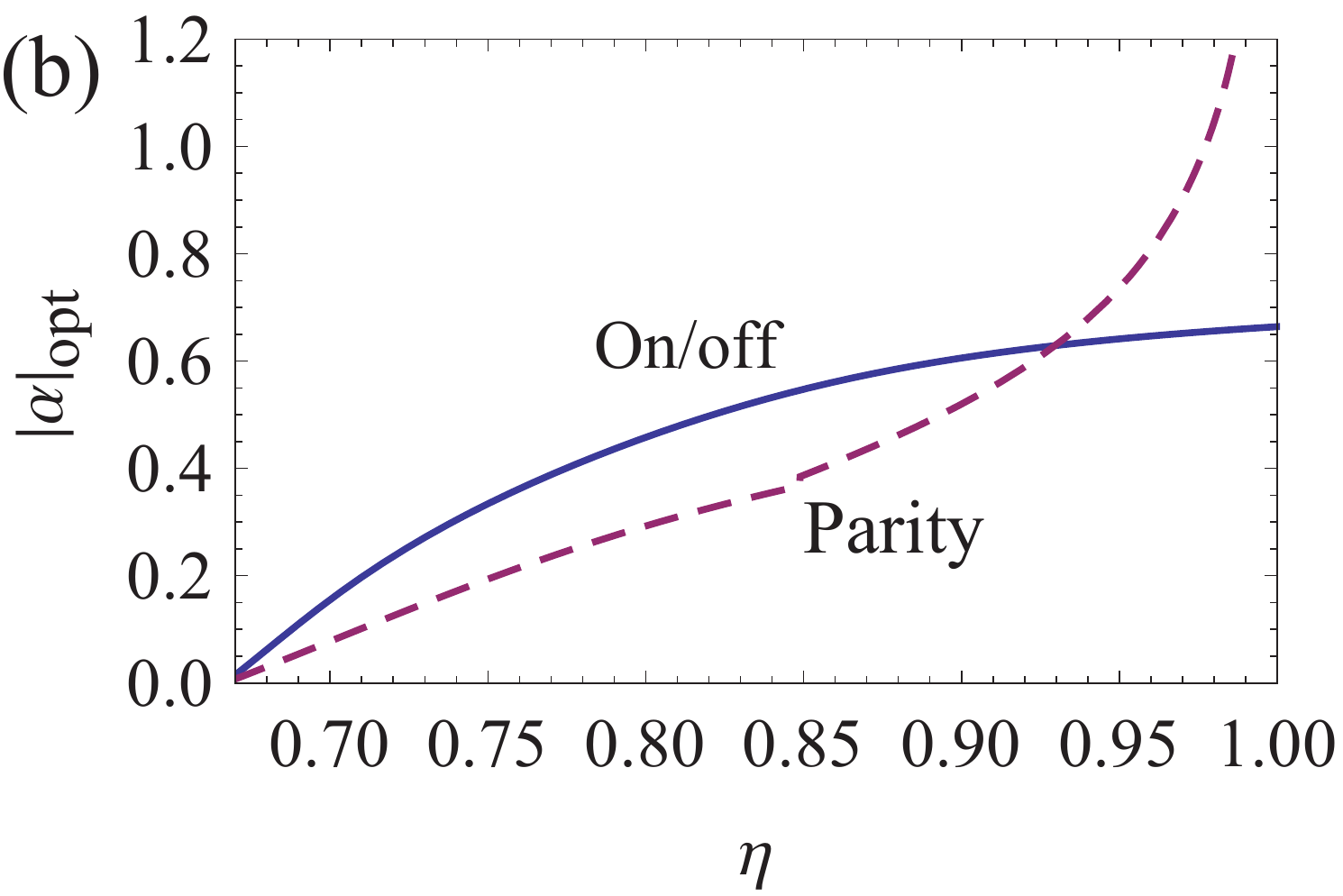}
	\caption{\small (Color online) (a) Comparison of Bell violation $|B|_\mathrm{max}$ (b) and optimizing coherent amplitude $|\alpha|_\mathrm{opt}$ between on/off(solid line) and parity(dashed line) measurements assuming symmetric detector efficiency $\eta$. On/off measurements give higher Bell value than parity measurements for $\eta<0.9868$.}
	\label{CompareB}
\end{figure}

We first assume the same detection efficiency $\eta_A = \eta_B=\eta$ on the polarization and the coherent field measurements. 
Figure~\ref{Contourna} shows that the degree of Bell violation and the optimizing coherent amplitudes for both the measurement schemes decrease when the detector efficiency $\eta$ decreases.
For example, with $\eta=0.8$, on/off measurements gives the maximum Bell violation of $|B|^\mathrm{on/off}_\mathrm{max} \approx 2.091$ at $|\alpha| \approx 0.458$, while parity measurements gives $|B|^\mathrm{parity}_\mathrm{max} \approx 2.035$ at $|\alpha| \approx 0.293$. If detector efficiency becomes $\eta=0.7$, the maximum Bell value and the optimizing coherent amplitude decrease to $|B|^\mathrm{on/off}_\mathrm{max} \approx 2.0022$ ($|\alpha| \approx 0.155$) and $|B|^\mathrm{parity}_\mathrm{max} \approx 2.0006$ ($|\alpha| \approx 0.078$) for each the measurement scheme. 
Figure \ref{coherentscaleO} and \ref{coherentscaleP} reveal that there is a trade-off between the degree of Bell violation and the detector efficiency  threshold by using different coherent amplitudes $|\alpha|$. Employing a low coherent amplitude demands low detection efficiency in order to see Bell inequality violation but the degree of the violation would be small.
Also we note from Fig. \ref{CompareB}(a) that with symmetric detector efficiency $\eta$ lower than $98.68\%$, on/off measurements provides higher Bell violation than parity measurements.

We numerically find that the Bell-CHSH inequality violation occurs until the detection efficiency reaches to 67\% for both the measurement schemes as presented in Fig. \ref{Contourna}. This value of the detection efficiency is lower than 82.8\% obtained by employing maximally entangled states and similar with the threshold efficiency for the Bell's inequality test using non-maximally entangled states \cite{PhysRevA.47.R747}.
We note that the maximum Bell violation occurs at $\left| \alpha \right| < 0.664$ for on/off measurements
 of any symmetric detector efficiency higher than the threshold efficiency 67\%. Similarly, the optimizing coherent amplitude for parity measurements is within the range $|\alpha| < 1.0$  when the detector efficiency is between the threshold (67\%) and 97.7\% (see Fig.~\ref{CompareB}(b)).

\begin{figure}[t]
	\centering
	\includegraphics[width=4.2cm]{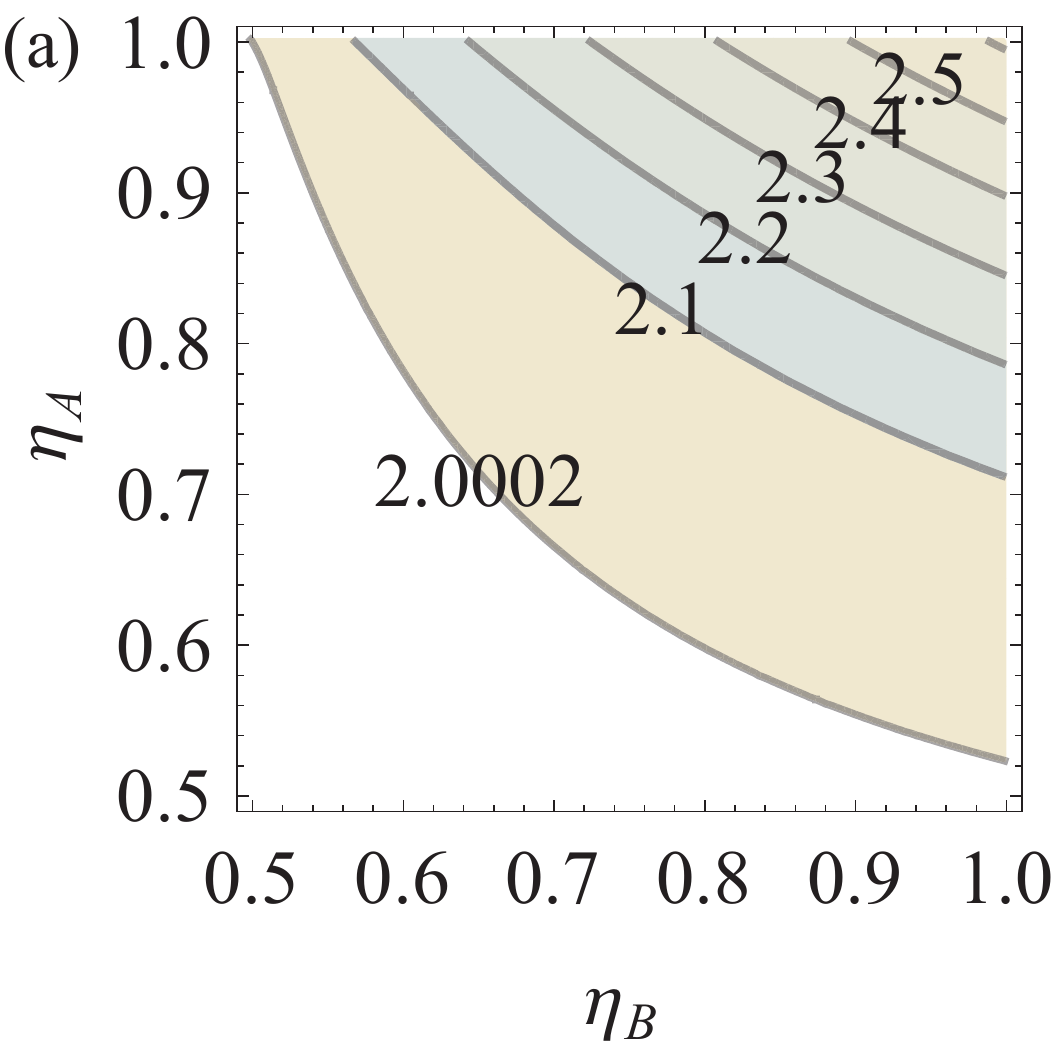} 
	\includegraphics[width=4.2cm]{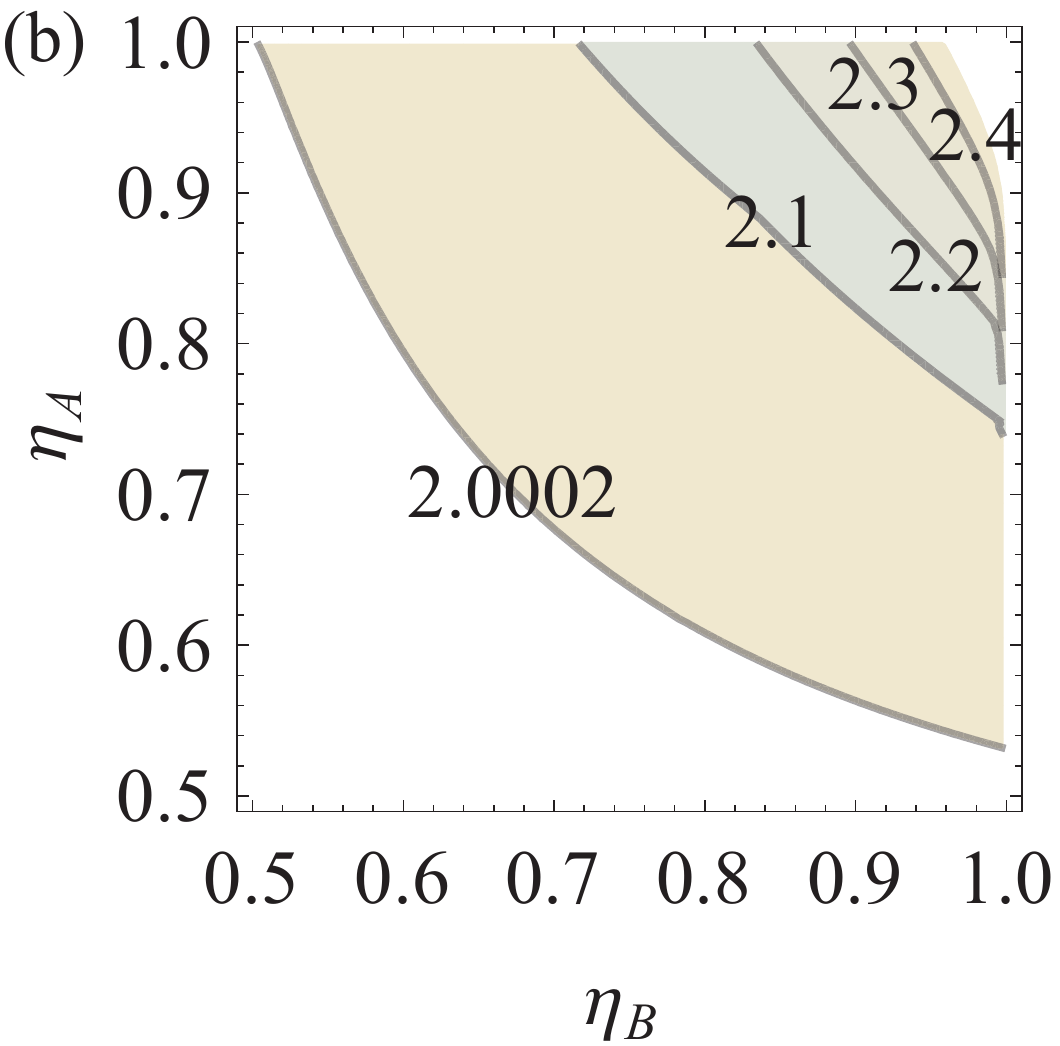}
	\caption{\small (Color online) Maximized Bell-CHSH value
	as a function of detection efficiencies for (a) on/off and (b) parity measurement schemes. Axis labels $\eta_A$ and $\eta_B$ refer to polarization (single photon) and coherent field measurement efficiencies, respectively. The optimizing coherent amplitudes were taken for each detection efficiency.}
	\label{Comparepq}
\end{figure}

In real experiments, actual values of the detection efficiency for the two separate local measurements may be different.
This realistic situation could be studied by taking the effective joint measurement defined by Eq.~(\ref{Eeff}) with local
detection efficiencies $\eta_A$ and $\eta_B$.
We plot the numerically optimized Bell function together with threshold regions for each measurement scheme in Fig.~\ref{Comparepq}.
The optimizing conditions have been found through nontrivial calculations as detailed in Appendix. 
As presented in Fig.~\ref{CompareB0}, the on/off measurement scheme provides higher violation of the Bell-CHSH inequality than the parity measurement scheme for most values of the detector efficiency.
Only when the coherent field detection efficiency $\eta_B$ is close to 1, parity measurements give higher Bell violation.
Figure~\ref{Comparepq} shows that the conditions for the Bell-CHSH inequality to be violated are similar for two different measurement schemes. This can be attributed to the facts that the probability distributions for two different measurement schemes ({\it i.e.} ``click" vs ``no-click" for the on/off scheme and ``odd" vs ``even" or the parity one) are similar for low coherent amplitudes and  low detection efficiency leads to the same optimizing conditions for both the measurement schemes.

\begin{figure}[t]
	\centering
	\includegraphics[width=4.5cm]{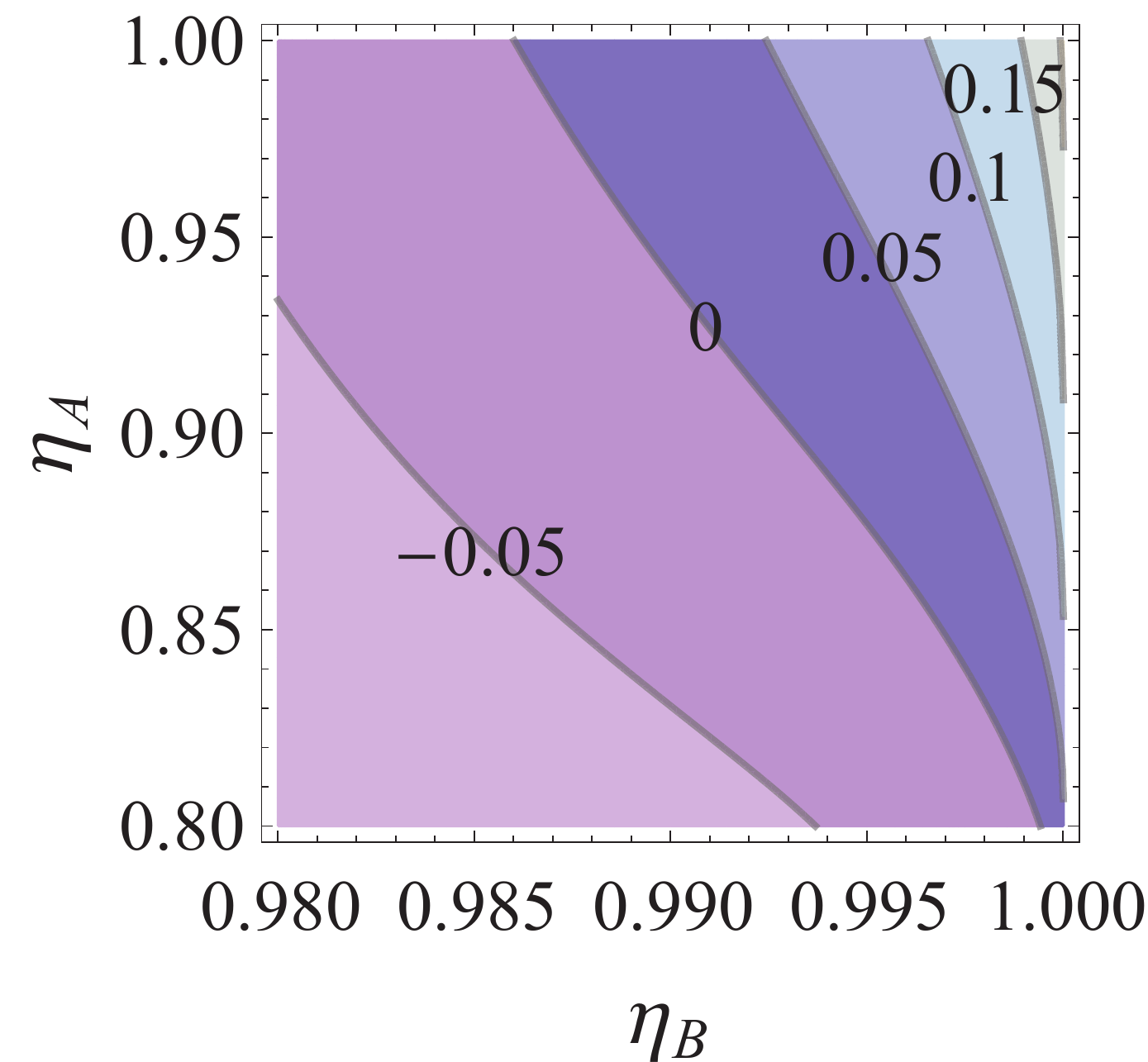}
	\caption{\small (Color online) Subtracted value, $|B|^\mathrm{Parity}_\mathrm{max}-|B|^\mathrm{On/off}_\mathrm{max}$, values for different detection efficiencies $\eta_A$ and $\eta_B$. Each $|B|_\mathrm{max}$ is obtained by taking its optimizing coherent amplitude.}
	\label{CompareB0}
\end{figure}

\section{Remarks}
We have studied Bell inequality test with hybrid entanglement between polarization of a single photon and a coherent state field. 
We have investigated two different kinds of measurements, on/off and photon number parity measurement, on the coherent field to find Bell violations with optimizing conditions with of perfect and realistic  detectors.
With perfect detectors, on/off measurements give the maximum Bell violation of $\approx 2.61$ at $\alpha \approx 0.664$, while parity measurements give the violation approaching Cirel'son's bound $(2\sqrt{2})$ for large values of the coherent amplitude ($\alpha \gg 1$).

 In order to see the Bell-CHSH inequality violation without the detection loophole, the detector efficiency $\eta > 67\%$ is required for both on/off and parity measurement schemes. It is important to note that small coherent amplitudes for hybrid entanglement are needed to obtain the low required efficiency while there is a trade-off between the threshold efficiency and the degree of Bell violation in terms of the coherent amplitudes. Nevertheless, a coherent amplitude of $|\alpha| < 1.0$ is sufficient to obtain the maximum Bell violation for the most cases of the detection efficiency.
Comparing two different measurement schemes, we have found that on/off measurements provide higher violation of Bell inequality than parity measurements under realistic conditions ($\eta < 98.68\% $), although the violation does not reach Cirel'son's bound. However, the threshold values of detection efficiency to violate Bell inequality are similar between both the measurement schemes.

Our results may be used to experimentally explore loophole-free Bell inequality tests.
Required detection efficiency for a loophole-free Bell test is within reach of current technology \cite{lita2008counting, hadfield2009single, marsili2013detecting}.
The generation of hybrid entanglement is a challenging task since it requires a clean cross-Kerr
nonlinearity, 
while efforts are being made to obtain high fidelity cross-Kerr interactions \cite{PhysRevA.83.053826, mahdi2012memory, PhysRevA.87.042325}.
It is also possible, in principle, to approximately generate arbitrary multimode entangled states using 
single-photon sources, coherent states, and single-photon detectors
\cite{JF2003}.
In this context, a possible attempt for the generation of hybrid entanglement  is to 
explore combinations of experimentally available photon addition and subtraction techniques 
\cite{AZ2004,JW2004,VP2007,MS2008,KimPRL2008,ZavattaPRL2009}
as investigated for the generation of some exotic quantum states
\cite{Marek2008,CW2012,SY2012}.

It is interesting to note that coherent states are considered most classical among all pure states
while single photons are typical microscopic quantum systems.
In this sense, Bell inequality tests using the optical hybrid states may reveal a significant feature of
nonlocality  between quantum and classical systems. 
It will be an interesting future work to explore quantum nonlocality with optical hybrid entanglement
using ``classical'' measurements \cite{classical1,classical2}.

\section*{ACKNOWLEDGMENTS}
This work was supported by the National Research Foundation of Korea(NRF) grant funded by the Korea government
(MSIP) (No. 2010-0018295) and by the Center for Theoretical Physics at Seoul National University.
H.K. was supported by NRF grant funded by the Korean Government (Global Ph.D. Fellowship Program No. 2012-003435).
\appendix*
\section{Optimizing conditions for imperfect detector efficiency}

\subsection{On/off measurements}
We have numerically found maximum Bell values and corresponding optimizing conditions
\cite{numerical}.
After numerical trials, we find that it is sufficient to 
take real values of the unitary parameters, $\xi$ and $\beta$, in order to obtain those
maximum Bell values for on/off measurements.
Under this condition, Eqs.~(\ref{Oeff1}) and (\ref{Oeff2}) can be represented in terms of $\theta$ and $|\beta|$ as

\small
\begin{equation}
\left \{
\begin{array}{l}
\left<\hat{O}_A \otimes \hat{O}^\mathrm{On/off}_{B,\mathrm{eff}} \right> = \mp2\cos\theta e^{-\eta_B\left(|\alpha|^2+|\beta|^2\right)} \sinh \left( 2\eta_B|\alpha||\beta| \right) \\ \qquad\qquad\qquad\qquad - \sin\theta e^{-2|\alpha|^2} + 2\sin\theta e^{-\left(2-\eta_B\right) |\alpha|^2 - \eta_B |\beta|^2}, \\
\mathrm{Tr}_B\left[\hat{O}^\mathrm{On/off}_{B,\mathrm{eff}} \rho_B\right] = 2 e^{-\eta_B\left(|\alpha|^2+|\beta|^2\right)} \cosh\left(2\eta_B|\alpha||\beta|\right) -1,
\end{array}
\right.
\label{OEeff}
\end{equation}
\normalsize
where $\mp$ corresponds to negative/positive $\beta$.
The Bell function $B$ can be constructed using Eqs.~(\ref{B}) and (\ref{Eeff}). In oder to find optimizing values, we take derivatives of $B$ to be zero  with respect to each parameter  as
\begin{equation}
\left\{ \frac{\partial B}{\partial \theta_1}, \frac{\partial B}{\partial \theta_2}, \frac{\partial B}{\partial \beta_1}, \frac{\partial B}{\partial \beta_2} \right\} = 0,
\label{eq:deriv}
\end{equation}
which leads to a set of equations
\small
\begin{equation}
\left\{
\begin{array}{l}
\tan\theta_1 = \frac{e^{-2\left(1- \eta_B \right)|\alpha|^2} \left( e^{- \eta_B |\alpha|^2} - e^{- \eta_B |\beta_1|^2} - e^{- \eta_B |\beta_2|^2} \right)}{e^{- \eta_B |\beta_1|^2}\sinh(2 \eta_B |\alpha||\beta_1|) - e^{- \eta_B |\beta_2|^2}\sinh(2 \eta_B |\alpha||\beta_2|)}, \\\\
\tan\theta_2 = -\frac{e^{-2\left(1- \eta_B \right)|\alpha|^2} \left(e^{- \eta_B |\beta_1|^2} - e^{- \eta_B |\beta_2|^2} \right)}{e^{- \eta_B |\beta_1|^2}\sinh(2 \eta_B |\alpha||\beta_1|) + e^{- \eta_B |\beta_2|^2}\sinh(2 \eta_B |\alpha||\beta_2|)},\\\\
\left(|\beta_1| \sinh(2 \eta_B |\alpha||\beta_1|) -|\alpha|\cosh(2 \eta_B |\alpha||\beta_1|) \right) \left( \cos\theta_1 - \cos\theta_2 \right) \\
\quad\qquad\qquad\qquad\qquad - e^{-2(1-\eta_B)|\alpha|^2} |\beta_1| \left( \sin\theta_1 - \sin\theta_2 \right) =0 ,\\\\
\left[ \left(-|\beta_2| \sinh(2\eta_B|\alpha||\beta_2|) + |\alpha|\cosh(2\eta_B|\alpha||\beta_2|) \right) \left( \cos\theta_1 + \cos\theta_2 \right) \right.\\
\qquad\qquad \left. - e^{-2(1-\eta_B)|\alpha|^2} |\beta_2| \left( \sin\theta_1 + \sin\theta_2 \right) \right] \\
  + \frac{2\left(1-\eta_A\right)}{\eta_A} \left( -|\beta_2| \cosh(2\eta_B|\alpha||\beta_2|) + |\alpha| \sinh(2\eta_B|\alpha||\beta_2|) \right) = 0.
\end{array}
\right.
\label{OOpteff}
\end{equation}
\normalsize

\subsection{Parity measurements}

\begin{figure}[b]
	\centering
	\includegraphics[width=6.cm]{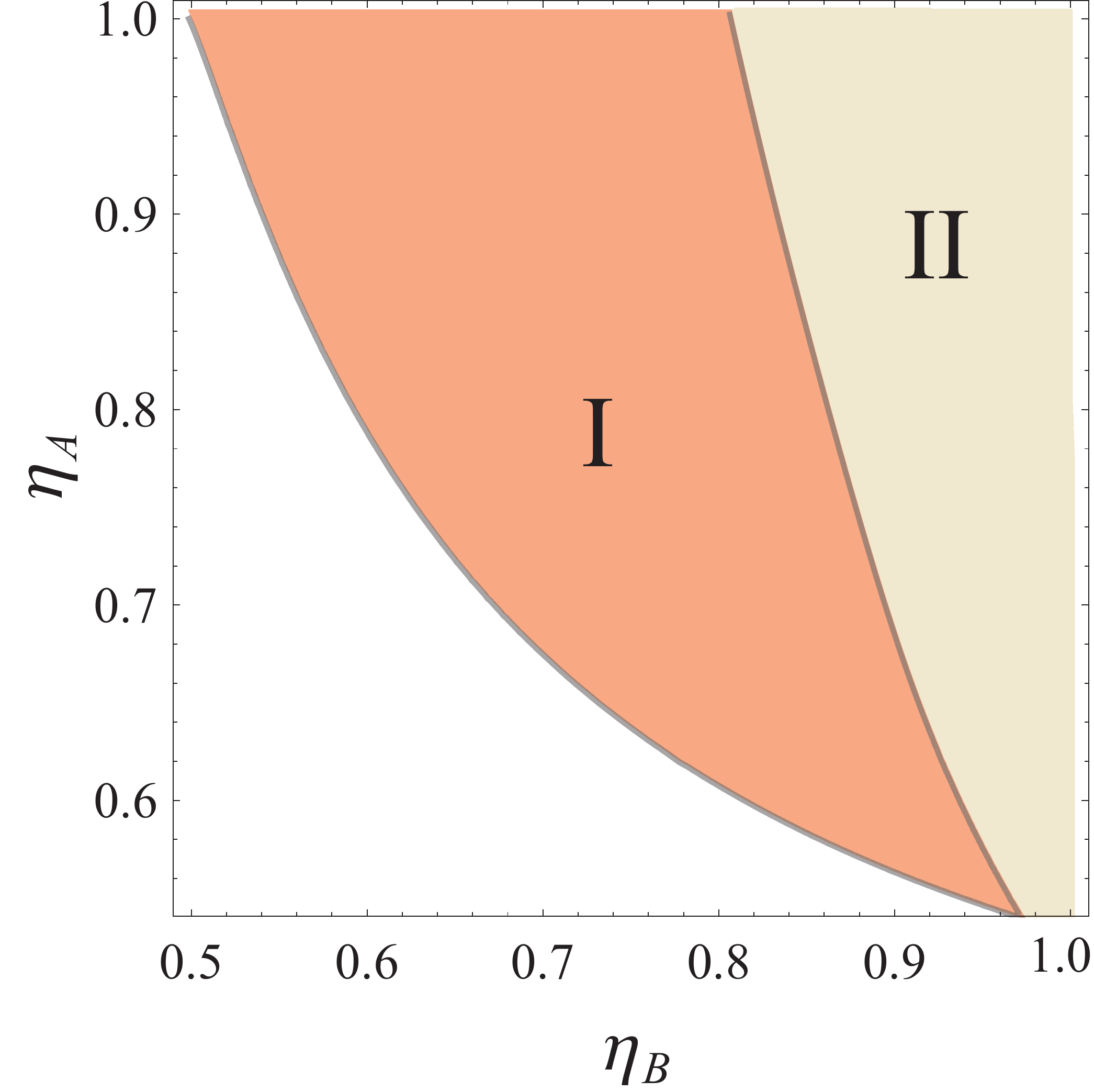}
	\caption{\small (Color online) Optimizing conditions for parity measurements with detection efficiencies
	$\eta_A$ and $\eta_B$.
	Shaded areas I and II are regions where the Bell-CHSH inequality is violated. Region I has optimizing conditions with real $\xi$ and $\beta$, while region II has optimizing conditions with $|\xi|={\pi}/{4}$ and pure imaginary $\beta$.}
	\label{ParityCondition}
\end{figure}

The scheme based on parity measurements undergoes two different optimizing conditions subject to
detection efficiencies $\eta_A$ and $\eta_B$. 
Region I in Fig.~\ref{ParityCondition} corresponds to optimizing conditions for low efficiency detectors, where real values of $\xi$ and $\beta$ are taken. We apply this condition to Eqs.~(\ref{Peff1}) and (\ref{Peff2}) and obtain the expectation value as
\begin{equation}
\begin{array}{l}
E^\mathrm{Parity,low}_\mathrm{eff} = \eta_A \left[ \mp2\cos\theta e^{-2\eta_B\left(|\alpha|^2+|\beta|^2\right)} \sinh \left( 4\eta_B|\alpha||\beta| \right) \right. \\
\qquad\qquad\qquad\qquad\qquad\qquad \left. + \sin\theta e^{-2\left(1-\eta_B\right) |\alpha|^2 - 2\eta_B |\beta|^2} \right] \\
\qquad\qquad\qquad + (1-\eta_A) e^{-2\eta_B\left(|\alpha|^2+|\beta|^2\right)} \cosh\left( 4\eta_B|\alpha||\beta|\right),
\end{array}
\end{equation}

\noindent where $\theta = -2\xi$ and $\mp$ corresponds to  negative/positive $\beta$. Using Eq.~(\ref{eq:deriv}), we
obtain a set of equations,

\small
\begin{equation}
\left\{
\begin{array}{l}
\tan\theta_1 = - \frac{e^{-2\left(1-2\eta_B\right)|\alpha|^2} \left(e^{-2\eta_B|\beta_1|^2} + e^{-2 \eta_B |\beta_2|^2} \right)}{e^{-2\eta_B|\beta_1|^2}\sinh(4\eta_B|\alpha||\beta_1|) - e^{-2\eta_B|\beta_2|^2}\sinh(4\eta_B|\alpha||\beta_2|)}, \\\\
\tan\theta_1 = - \frac{e^{-2\left(1-2\eta_B\right)|\alpha|^2} \left(e^{-2\eta_B|\beta_1|^2} - e^{-2\eta_B|\beta_2|^2} \right)}{e^{-2\eta_B|\beta_1|^2}\sinh(4\eta_B|\alpha||\beta_1|) + e^{-2\eta_B|\beta_2|^2}\sinh(4\eta_B|\alpha||\beta_2|)}, \\\\
\left(|\beta_1| \sinh(4\eta_B|\alpha||\beta_1|) - |\alpha| \cosh(4\eta_B|\alpha||\beta_1|) \right) \left( \cos\theta_1 - \cos\theta_2 \right) \\
\quad\qquad\qquad\qquad\qquad - e^{-2(1-2\eta_B)|\alpha|^2} |\beta_1| \left( \sin\theta_1 - \sin\theta_2 \right) =0, \\\\
\left[ \left(-|\beta_2| \sinh(4\eta_B|\alpha||\beta_2|) + |\alpha|\cosh(4\eta_B|\alpha||\beta_2|) \right) \left( \cos\theta_1 + \cos\theta_2 \right) \right.\\
\qquad\qquad \left. - e^{-2(1-2\eta_B)|\alpha|^2} |\beta_2| \left( \sin\theta_1 + \sin\theta_2 \right) \right] \\
+\frac{2\left(1-\eta_A\right)}{\eta_A} \left( -|\beta_2| \cosh(4\eta_B|\alpha||\beta_2|) + |\alpha| \sinh(4\eta_B|\alpha||\beta_2|) \right) = 0,
\end{array}
\right.
\end{equation}
\normalsize
and find the optimizing conditions.

Optimization for the high detection efficiencies (region II in Fig.~\ref{ParityCondition}) can be obtained by taking $|\xi_1| = |\xi_2| = {\pi}/{4}$ and $\beta$ as pure imaginary number. The expectation value then becomes

\begin{equation}
\begin{array}{l}
E^\mathrm{Parity,high}_\mathrm{eff} = \eta_A ~ e^{-2\left(1-\eta_B\right) |\alpha|^2 - 2\eta_B |\beta|^2} \cos \left( 4 \eta_B |\alpha||\beta| \pm \phi \right) \\
\qquad\qquad\qquad + (1-\eta_A) e^{-2\eta_B\left(|\alpha|^2+|\beta|^2\right)} \cosh \left( 4\eta_B |\alpha| |\beta|\right),
\end{array}
\end{equation}
where $\phi$ is a phase factor of $\xi$ and $\pm$ corresponds with negative/positive $-i\beta$. In this case, we take similar steps with Eq.~(\ref{eq:deriv}) but using a set of parameters of $\{\phi_1$, $\phi_2$, $|\beta_1|$,$|\beta_2|\}$, and find optimizing conditions by solving a set of equations:
\small
\begin{equation}
\left\{
\begin{array}{l}
e^{-2 \eta_B |\beta_1|^2} \sin \left( 4\eta_B|\alpha||\beta_1| + \phi_1 \right) \\
\qquad\qquad\qquad\qquad\qquad = e^{-2\eta_B|\beta_2|^2} \sin\left(4\eta_B|\alpha||\beta_2| - \phi_1\right), \\\\
e^{-2\eta_B|\beta_1|^2} \sin \left( 4\eta_B|\alpha||\beta_1| + \phi_2 \right) \\
\qquad\qquad\qquad\qquad\qquad = - e^{-2\eta_B|\beta_2|^2} \sin\left(4\eta_B|\alpha||\beta_2| - \phi_2\right), \\\\
|\beta_1| \left( \cos(4\eta_B|\alpha||\beta_1| + \phi_1) - \cos(4\eta_B|\alpha||\beta_1|+ \phi_2)\right) \\
\qquad+ |\alpha| \left( \sin(4\eta_B|\alpha||\beta_1|+\phi_1) - \sin(4\eta_B|\alpha||\beta_1|+\phi_2)\right) = 0, \\\\
|\beta_2| \left( \cos(4\eta_B|\alpha||\beta_2| - \phi_1) + \cos(4\eta_B|\alpha||\beta_2|- \phi_2)\right) \\
\qquad + |\alpha| \left( \sin(4\eta_B|\alpha||\beta_2|-\phi_1) + \sin(4\eta_B|\alpha||\beta_2|-\phi_2)\right) \\
~~~~~~\qquad\qquad\qquad\qquad\qquad + \frac{2(1-\eta_A)}{\eta_A} |\beta_2| e^{2(1-2\eta_B)|\alpha|^2} = 0.
\end{array}
\right.
\end{equation}

\end{document}